# Towards Accurate and Efficient Sorting of Retired Lithium-ion Batteries: A Data Driven Based Electrode Aging Assessment Approach

Ruohan Guo[*,a], Feng Wang[a], Cungang Hu[b], Weixiang Shen[a]

[a]School of Science, Computing and Engineering Technologies, Swinburne University of Technology, Hawthorn, Victoria 3122, Australia

[b]School of Electrical Engineering and Automation, Anhui University, Hefei, Anhui 230601, China

**Abstract:** Retired batteries (RBs) for second-life applications offer promising economic and environmental benefits. However, accurate and efficient sorting of RBs with discrepant characteristics persists as a pressing challenge. In this study, we introduce a data driven based electrode aging assessment approach to address this concern. To this end, a number of 15 feature points are extracted from battery open circuit voltage (OCV) curves to capture their characteristics at different levels of aging, and a convolutional neural network with an optimized structure and minimized input size is established to relocate the relative positions of these OCV feature points. Next, a rapid estimation algorithm is proposed to identify the three electrode aging parameters (EAPs) which best reconstruct the 15 OCV feature points over the entire usable capacity range. Utilizing the three EAPs as sorting indices, we employ an adaptive affinity propagation algorithm to cluster RBs without the need for pre-determining the clustering number. Unlike conventional sorting methods based solely on battery capacity, the proposed method provides profound insights into electrode aging behaviors, minimizes the need for constant-current charging data, and supports module/pack-level tests for the simultaneous processing of high volumes of RBs.

[*] Corresponding author (Email: rguo@swin.edu.au)

1. **Introduction**

As the global transition towards electrified transportation gains momentum, lithium-ion batteries (LIBs) have emerged as the leading power source for electric vehicles (EVs) [1–3]. This preference stems from their impressive discharging/charging capabilities, cost-effective production, and extended service life [4,5]. However, LIBs demonstrate various aspects of degradation over lifetime, when their remaining capacities reach 80% of the nominal capacity or internal resistances (IRs) increase to 200% of their initial states, this usually marks the end of life (EOL) of an EV battery pack [6–8]. Interestingly, even though all in-pack battery cells undergo the same usage histories, their aging behaviors can vary. Such variation arises from manufacturing inconsistencies and differences in cell placement, temperature, and state of charge (SOC), resulting in significant heterogeneity within a pack [9–11]. It has been reported by many previous studies [12,13] that over 70% of in-pack battery cells may still reserve more than 80% of the nominal capacity when an EV battery pack reaches its EOL.

Direct recycling of those retired batteries (RBs) with high remaining capacities could exacerbate environmental pollutions on air/water/land, waste substantial energy and resources, and inflate recycling costs [14]. Clearly, it is an imprudent decision from both environmental and economic standpoints [15]. In recent years, reusing RBs in stationary storage applications has gained traction, offering significant environmental and economic advantages. This can not only extend battery service life but also optimize resource utilization and open up new revenue opportunities. However, a significant challenge arises from the absence of standardized procedures and guidelines in the battery recycling industry, particularly when sorting high volumes of RBs with discrepant characteristics.

Currently, most sorting techniques rely on carefully selected sorting indices. For instance, a range of battery external characteristics with high accessibility, such as shape [16], weight [17], self-discharge rate [18], thermal behavior [19], available capacity [20], and IR [21,22], have been extracted to serve as sorting indices, and their effectiveness has been extensively examined in the literature. To facilitate sorting performance and guarantee operational safety in battery second life, multi-index sorting methods have received much attention from both industries and academies. In [23], available capacity, pulse discharging voltage, charge transfer resistance, and diffusion coefficient were introduced to jointly sort RBs in good performance. Jiang et al. [24] took into account available capacity, IRs at each 10% SOC, and peak heights of incremental capacity (IC) curves as multi-indices and utilized a fuzzy K-means clustering (KMC) algorithm to improve sorting accuracy. Through an in-depth exploitation of the IC curves, a set of seven indices, namely two peak heights with the associated voltages and the slope of both sides of one peak with its covered regional area, was identified with certain correlations with battery capacity decay [25]. In light of this, a sorting method was proposed by taking advantages of both modified KMC algorithm and T-test to alleviate the impact of inappropriate initial centroid selections and make RBs more concentrated in their respective clusters. In [26], a few characteristic voltages and IRs were manually selected from constant current (CC) charging curves, and a weighted KMC algorithm was applied to sort RBs considering different application scenarios in their second life. Their subsequent study in [27] investigated the viability of leveraging electrochemical impedance spectrum (EIS) technique (spanning from 0.01 Hz to 1 kHz) and distribution of relaxation time analysis in battery sorting. They characterized the contact resistance, solid-electrolyte interface (SEI) resistance, charge transfer resistance, and diffusion resistance, and developed a weighted Gaussian mixture model based clustering method to factor in the contributions of these indices to different degradation patterns.

Besides those methods based on manual extractions of sorting indices, automated feature engineering techniques were also employed in battery sorting. For instance, Zhou et al. [28] picked up three sorting indices by performing the principal component analysis on the experimental results of pulse discharging and charging and proposed an improved bisection KMC algorithm to promote the sorting robustness against noise disturbances. Lin et al. [29] devised an automated sorting process, where the CC charging data was directly transformed into a Markov transition field, followed by a swin transformer for clustering. A two-stage sorting scheme was put forward in [30]; in the first stage, a density based spatial clustering algorithm was used to eliminate abnormal RBs in terms of available capacity and temperature rise; in the second stage, a mass of discharging data was fed into a T-distributed stochastic neighbor embedding algorithm, outputting useful sorting indices with reduced dimensions, and a KMC algorithm was adopted for ultimate clustering.

While incorporating more sorting indices generally enhances sorting performance, it comes with associated drawbacks, notably high experimental cost and low efficiency. This is primarily because many of the sorting indices are obtained from cell/module-level experiments, such as CC tests, pulse tests, and EIS tests. The labor-intensive battery disassembly and the time-consuming discharging/charging experiments will inevitably slow down the entire sorting progress. In addition, it is uninterpretable for a majority of sorting indices regarding their relationships with battery degradation, despite they vary with battery capacity or IR. As battery capacity decay or internal resistance increase is only a matter of outward manifestation of battery degradation, it is likely that different degradation patterns of two RBs give rise to similar external characteristics. Such possible coincidences can affect the performance of second-life batteries and risk their operational safety. It underlines the capabilities of sorting indices to reflect the internal characteristics of a battery. For instance, some researchers captured electrode morphological changes [31] or recalibrated electrochemical parameters [32] to intuitively sort RBs with variable degrees of degradation, but it required high-demanding equipment with domain expertise for conducting experiments and data analysis, making it only feasible in laboratory scale.

Targeting the aforementioned research gaps, this study proposes an electrode aging assessment approach for sorting RBs, making the following four contributions:

1) Three novel electrode aging parameters (EAPs) are introduced to assess the electrode aging behavior of a battery. They have high accessibility by using only a few CC charging data and are capable of characterizing the effects of loss of lithium inventory (LLI) and loss of active material (LAM) which help explain the root cause of battery capacity decay and open circuit voltage (OCV) distortion.

2) A number of 15 feature points are extracted from battery OCV curves to capture their characteristics at different levels of aging, and a convolutional neural network (CNN) with an optimized structure and minimized input size is established to relocate the relative positions of these OCV feature points. On this basis, a rapid estimation algorithm is developed to identify the optimal EAPs which best reconstruct 15 OCV feature points over the entire usable capacity range.

3) Utilizing the three EAPs as sorting indices, we employ an adaptive affinity propagation (adAP) algorithm to cluster RBs, which unlocks the necessity of pre-determining the clustering number while addressing the issues of preference tuning, convergence, and oscillations during iterations.

4) The proposed electrode aging assessment approach brings several desirable attributes: (1) it provides profound insights into various electrode aging behaviors of RBs, contributing to sorting performance at an electrode level; (2) it minimizes the need for CC charging data, significantly reducing experimental effort; and (3) it supports module/pack-level tests for the simultaneous processing of high volumes of RBs, indicating great potential for industrial scale-up applications.

The remainder of this paper is structured below. The battery experiments are introduced in Section 2. Section 3 illustrates the proposed method for electrode aging assessment and battery sorting. Section 4 presents validation results and discussions. Conclusions are drawn in Section 5.

## 2. Battery experiments

### 2.1. Battery degradation dataset

The proposed method is developed based on the Oxford battery degradation dataset 1 [1]. This complete dataset contains the degradation data of eight Kokam pouch batteries, labelled as Cell #1~8, and their capacity decay trajectories throughout the whole lifetimes are depicted in Fig. 1. The Kokam pouch batteries take lithium cobalt/nickel cobalt oxide and graphite for the positive and negative electrode materials, respectively, and have a nominal capacity of 740 mAh. During aging tests, the batteries underwent CC charging at 2 C-rate and dynamic discharging under an EV driving profile to mimic real-world usage conditions. CC charging and discharging tests at 1 C-rate and 1/20 C-rate were conducted at every 100 cycles to calibrate the variations of battery available capacity and OCV curves over the whole battery lifetime. The voltage limits were set to 4.2 V (upper cut-off) and 2.7 V (lower cut-off), respectively. A Bio-Logic MPG-205 battery tester was employed to perform the test program and record experimental data with a sampling frequency of 1 Hz. The batteries were housed in a thermal chamber, maintaining a stable ambient temperature of 40 Celsius throughout all the experiments. The total discharging capacity at 1/20 C-rate is regarded as the available capacity of a battery at each aging level. Reference battery OCVs are acquired by taking the average of the voltage measurements under charging and discharging trajectories at 1/20 C-rate to offset the hysteresis effect. A Savitzky-Golay filter is applied to process these OCV data for mitigating the noise interference. Fig. 2 (a) demonstrates battery OCV and 1-C charging/discharging curves of Cell #1 after data pre-processing, and Fig. 2 (b) exhibits the reference battery OCV curves over the whole lifetime.

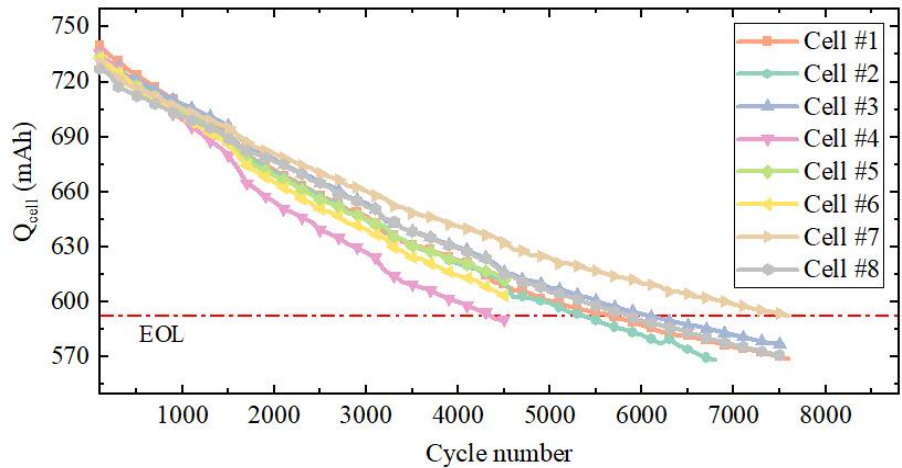

**Fig. 1.** Capacity decay trajectories of Cell #1~8.

A commonly used EOL standard for LIBs serving in EVs is adopted in this study, namely the available battery capacity reaches 80% of its nominal capacity. As can be seen from Fig. 1, the available capacities of some batteries have been under the EOL standard, making these degradation data well-suited to act as the overused RBs from EVs. Therefore, we separate the whole dataset into two portions. The first portion contains Cell #1~3, serving as the prior knowledge of this type of batteries while the latter one contains Cell #4~8, serving for validations of the proposed method.

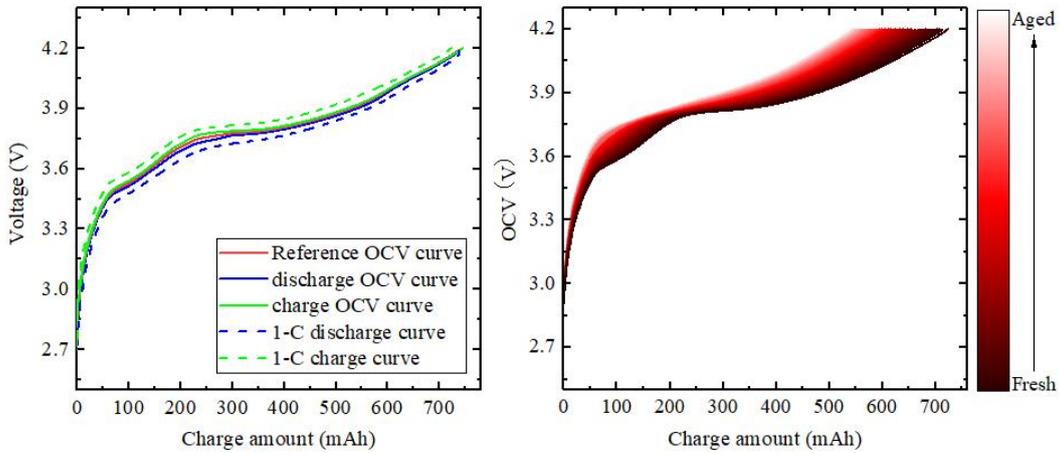

**Fig. 2.** Battery OCV and 1-C charging/discharging curves of Cell #1 (a) and the obtained reference battery OCV curves over the whole lifetime (b).

2.2. Electrode OCV model

Apart from the aforementioned battery experiments at a cell level, Dr Brikl also performed half-cell tests with the electrode material collected from a same-type Kokam pouch battery in [34]. This allows for the construction of electrode OCV models in Eq. (1) to accurately capture the electrode OCV-SOC relationships shown in Fig. 3.

$$\begin{cases} x_{PE}(E_{PE}) = \sum_{i=1}^{5} \dfrac{\Delta x_{PE,i}}{1+\exp\{(E_{PE}-E_{0,PE,i})\zeta_{PE,i}e/(k_BT)\}} \\ x_{NE}(E_{NE}) = \sum_{i=1}^{5} \dfrac{\Delta x_{NE,i}}{1+\exp\{(E_{NE}-E_{0,NE,i})\zeta_{NE,i}e/(k_BT)\}} \end{cases} \quad (1)$$

where $T$ is the absolute temperature in the unit of K, $k_B$ is the Boltzmann constant, and $e$ is the elementary charge. $x$ gives the ratio of intercalated lithium quantities to the total available sites in PE/NE, which can be deemed as electrode SOC. $E$ stands for the electrode OCV. The model parameters to be identified include the fraction of occupied sites $\Delta x$, the standard redox potential $E_0$, and a fitting parameter $\zeta$.

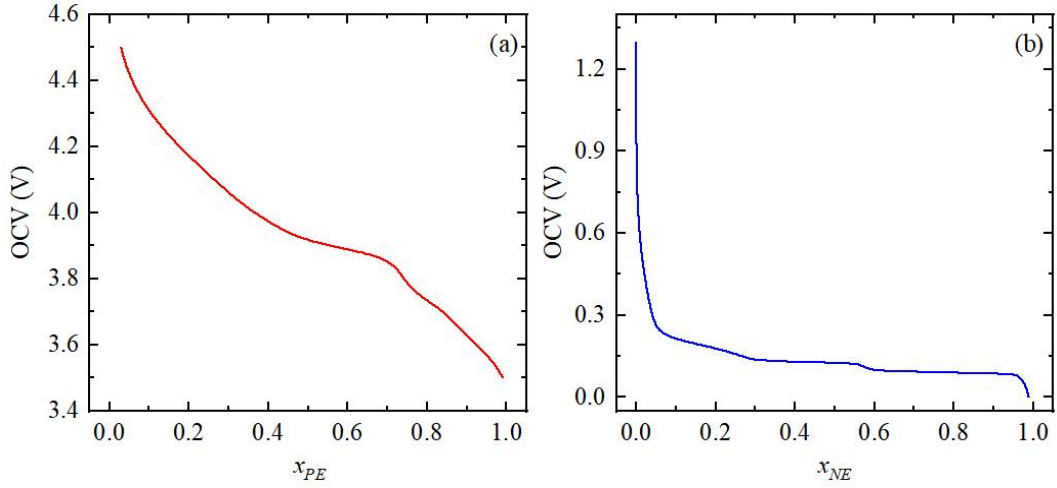

**Fig. 3.** OCV-SOC curves of PE (a) and NE (b).

**3. Electrode aging assessment approach for battery sorting**

3.1. Aging mechanism analysis

The previous study [35,36] has highlighted that the OCV-SOC correlations of positive electrode (PE) and negative electrode (NE) are barely affected by battery degradation. This inherent stability contributes to the understanding of battery aging mechanism, unveiling the root cause of capacity decay and OCV distortion for aged cells. To provide a visual representation of battery aging mechanism, we establish a coordinate system in Fig. 4, with the left start point of PE OCV curve serving as the origin on the x-axis. In this coordinate system, we depict the relative positions of electrode and battery OCV curves in one plot. It is observed that for a fresh cell the NE OCV curve is shifted to the right by a certain offset, attributed to the SEI formation during battery manufacturing [37]. This phenomenon results in an irreversible capacity loss at the beginning of battery lifetime. We denote this capacity offset as $Q_{\text{offset}}$. Given $Q_{\text{offset}}$ with the capacity of PE and NE (i.e., $Q_{PE}$ and $Q_{NE}$), we could uniquely determine a battery OCV curve over the entire usable capacity range by

$$V_{oc} = f_{PE}(x_{PE}) - f_{NE}(x_{NE}) \quad (2)$$

with

$$x_{PE/NE} = \frac{Q_{inserted}}{Q_{PE/NE}} \quad (3)$$

where $Q_{inserted}$ denotes the inserted lithium quantities at PE/NE. $f_{PE/NE}(x_{PE/NE})$ is described by electrode OCV models in Eq. (1).

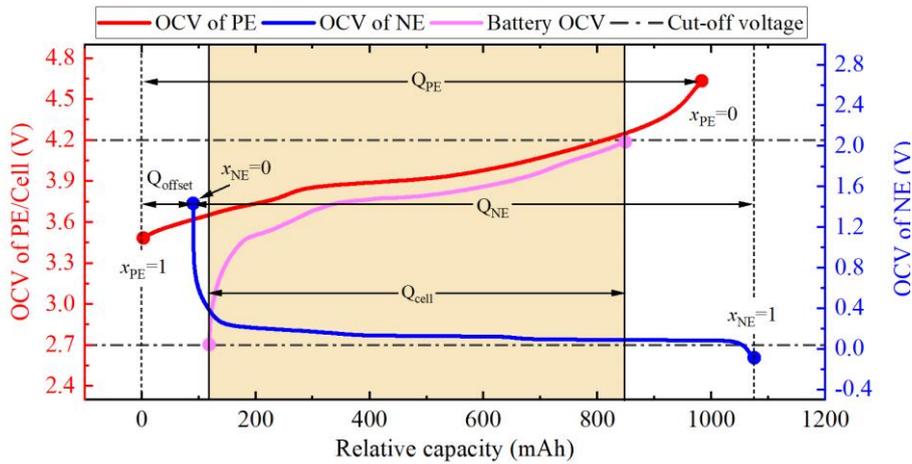

**Fig. 4.** Relationship between electrode and battery OCV curves of a fresh cell.

LLI and LAM represent two dominant factors contributing to capacity decay and OCV distortion during a battery degradation process [38]. Fig. 5 gives a detailed visualization of how LLI and LAM affect battery available capacity and OCV curve based on the proposed coordinate system. Referring to the top subplot of Fig. 5, LLI expands $Q_{offset}$ between the start points of PE and NE OCV curves on the left side. This phenomenon directly reduces the overlapping region between the two electrode OCV curves on the x-axis, thereby decreasing the available capacity of a battery. Additionally, an increasing $Q_{offset}$ also leads to an early drop of battery OCV curve in the middle SOC region. On the other hand, LAM exerts a more intricate influence on battery degradation from two aspects: (1) LAM of (de)lithiated PE, denoted as $LAM_{dePE}$ and $LAM_{liPE}$; and (2) LAM of (de)lithiated NE, denoted as $LAM_{deNE}$ and $LAM_{liNE}$. As demonstrated in the middle and bottom subplots of Fig. 5, $LAM_{liPE}$ and $LAM_{deNE}$ reduce the capacities of both PE and NE from right to left while $LAM_{dePE}$ and $LAM_{liNE}$ reduce the capacities from left to right. As a consequence, LAM narrows the usable capacity range of a battery, inevitably causing a capacity decay and an OCV distortion (e.g., sharper slope).

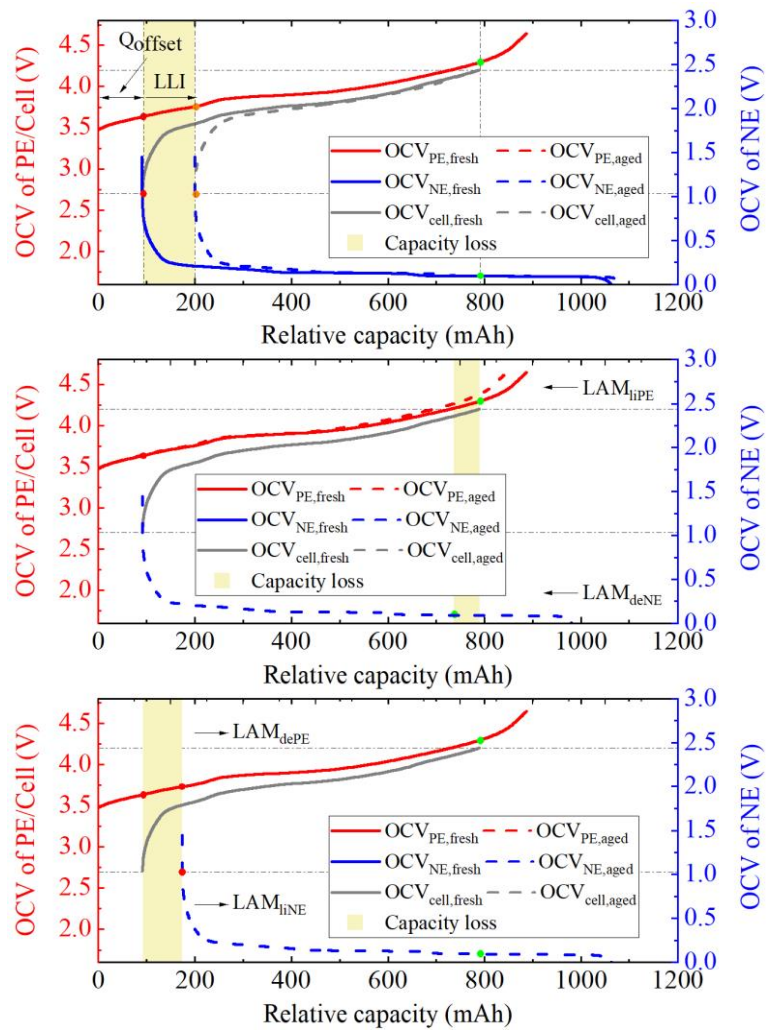

**Fig. 5.** LLI- and LAM-induced battery aging mechanism and associated effects on electrode and battery OCV curves.

According to the discussion above, the adverse effects of LLI and LAM on battery available capacity and OCV curve can be captured via $Q_{PE}$, $Q_{NE}$ and $Q_{offset}$. In light of this, we adopt them as three EAPs to evaluate different degradation patterns of RBs, subsequently achieving accurate and efficient sorting at an electrode level. The detailed EAPs estimation and clustering method will be introduced in the following subsections.

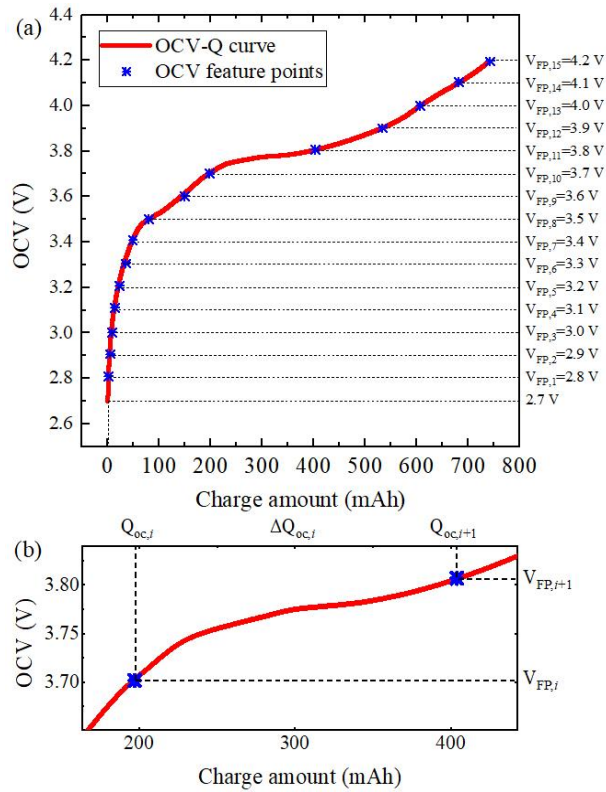

**Fig. 6.** Battery OCV feature points (a) and differential charge amount between two adjacent OCV feature points (b).

3.2. Battery OCV feature point selection and relocation

A battery OCV curve encompasses a mass of data points, making it a difficult and labor-intensive task to reconstruct the entire curve. To cope with this issue, we only pick up a few OCV feature points and use their relative positions to characterize the OCV shapes at various aging levels. Let the feature voltages be the values from 2.8 V to 4.2 V at 0.1 V steps (i.e., $V_{FP,1} \sim V_{FP,15}$), a total of 15 feature points, as shown in Fig.

6 (a), can be determined on a battery OCV curve, and their corresponding charge amounts (i.e., $Q_{FP,1} \sim Q_{FP,15}$) incur variable degrees of shifting with battery degradation. Since the charge amounts of the OCV feature points on the right have certain dependencies on those OCV feature points on the left, we further convert $Q_{FP,1} \sim Q_{FP,15}$ into differential charge amounts between two adjacent OCV feature points by

$$\begin{cases} \Delta Q_{FP,i} = Q_{FP,i} - \Delta Q_{FP,i-1}, & i \geq 2 \\ \Delta Q_{FP,i} = Q_{FP,i}, & i = 1 \end{cases} \quad (4)$$

where $Q_{FP,i}$ denotes the charge amount of the $i$-th OCV feature point and $\Delta Q_{FP,i}$ denotes the differential charge amount between the $i$-th and ($i$-1)-th OCV feature points (see Fig. 6 (b)). Indeed, the differential charge amounts reflect the relative positions of the OCV feature points, and their original charge amounts can be easily retrieved by

$$Q_{FP,i} = \sum_{j=1}^{i} \Delta Q_{FP,j} \quad (5)$$

Fig. 7 exemplifies $\Delta Q_{FP,1} \sim \Delta Q_{FP,15}$ of Cell #1 over the whole battery lifetime, and their diverse variation patterns indicate the capability of the OCV feature points in reflecting the battery OCV characteristics for aged cells. In this study, we estimate $\Delta Q_{FP,1} \sim \Delta Q_{FP,15}$ to relocate the relative positions of the OCV feature points based on partial IC segments. Battery IC curves are defined as the differentiated battery capacity in terms of voltage, which can be mathematically expressed as

$$IC = \frac{dQ_{cc}}{dV_t} \simeq \frac{\Delta Q_{cc}}{\Delta V_t} \quad (6)$$

where $Q_{cc}$ stands for the charge accumulation. $V_t$ denotes the battery terminal voltage, and $\Delta V_t$ gives the voltage sampling step, which is set to be 1 mV in this study. Fig. 8 presents the 1-C charging and associated IC curves of Cell #1 over the whole battery lifetime. As can be seen in Fig. 8 (d), almost all the peaks and valleys of IC curves are concentrated within [3.5 V, 4 V], indicating high sensitivity to battery degradation and rich aging information therein. Hence, we construct a CNN and pick up partial IC segments within [3.5 V, 4 V] as the model input for relocating the OCV feature points.

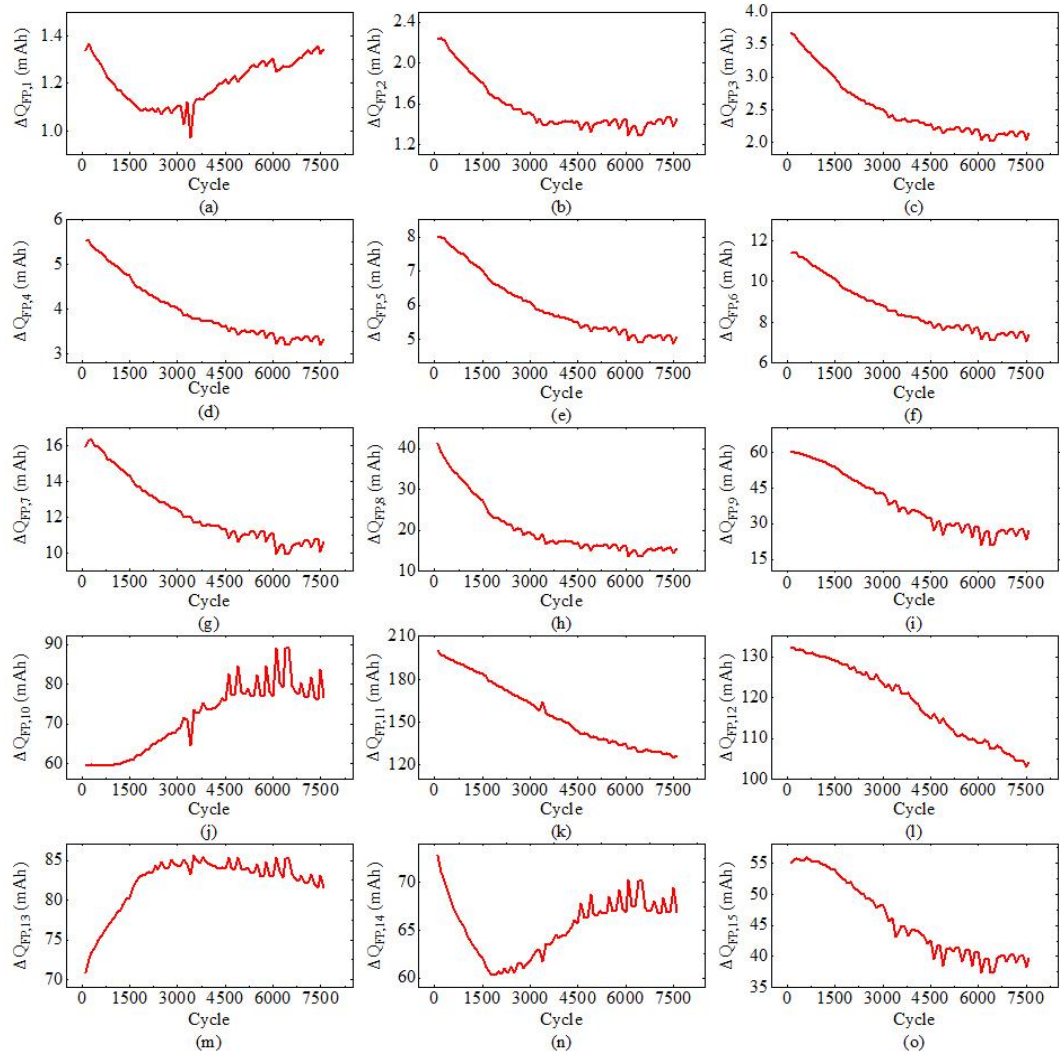

**Fig. 7.** Variation trajectories of differential charge amounts between two adjacent OCV feature points of Cell #1 over the whole battery lifetime.

The architecture of the proposed CNN is demonstrated in Fig. 9 (a), which constitutes a number of convolutional blocks, followed by several dense layers. Each of the convolutional blocks consists of a convolutional layer and a max pooling layer. As illustrated in Fig. 9 (b), a convolutional layer adopts a set of filters to implement convolution operations, extracting the underlying features from layer inputs while outputting the dot products. The filter weights and biases in a convolutional layer are utilized uniformly across the entire input sequence, which prove to be effective in feature extraction [39]. After a convolutional layer, a following max pooling layer serves to reduce data dimension, thereby alleviating the risk of overfittings. The relevant features extracted from the convolutional blocks are flattened and transmitted into dense layers to learn the inner relationships with the model outputs (i.e., $\Delta Q_{FP,1} \sim \Delta Q_{FP,15}$), as illustrated in Fig. 9 (c). The rectified linear unit activation function is employed in these dense layers to increase nonlinear interpretation capability. The degradation data from Cell #1~3 and Cell #4~8 are assigned as training set and testing set, respectively. To alleviate overfittings, 10% of the training data are randomly selected to form a validation set. Before training, all the data are normalized within a range of [-1,1] via a minmax function. The Adam algorithm is adopted as an optimizer in the training process to update the weights and biases for the CNN. The root mean square error (RMSE) is used to assess the model performance.

An appropriate selection of hyperparameters is crucial for improving the precision of a deep learning model. In addition, a lengthier partial IC segment usually benefits the feature preservation of battery aging information, leading to enhanced model performance. Nevertheless, it demands increased experimental efforts in battery testing and data acquisition, potentially reducing sorting efficiency, particularly for an industrial process. In this regard, optimizing the hyperparameters and input size of the proposed CNN can be treated as a multi-objective minimization problem, described by two potentially conflicting objective functions in Eq. (7).

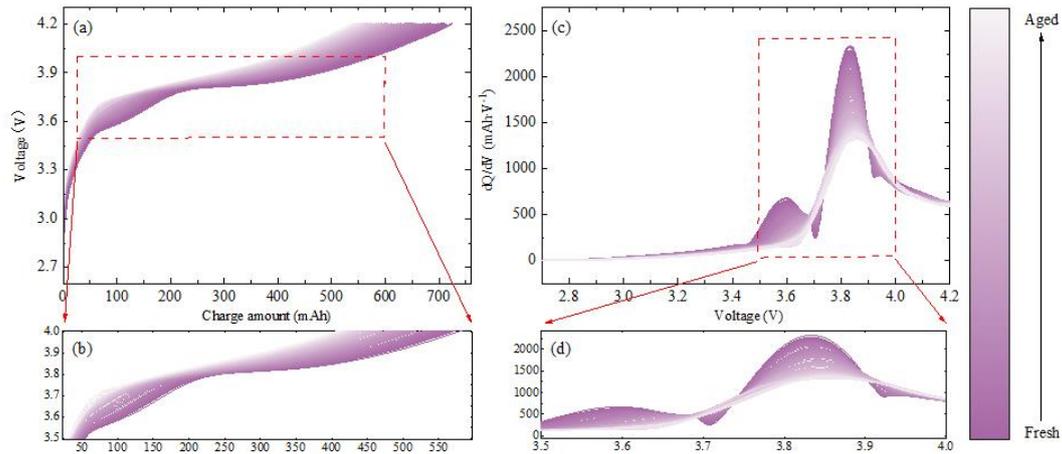

**Fig. 8.** 1-C charging curves and the associated IC curves of Cell #1 over the whole battery lifetime: (a) 1-C charging curves; (b) zoom-in view of (a) within [3.5 V, 4 V]; (c) IC curves; (d) zoom-in view of (c) within [3.5 V, 4 V].

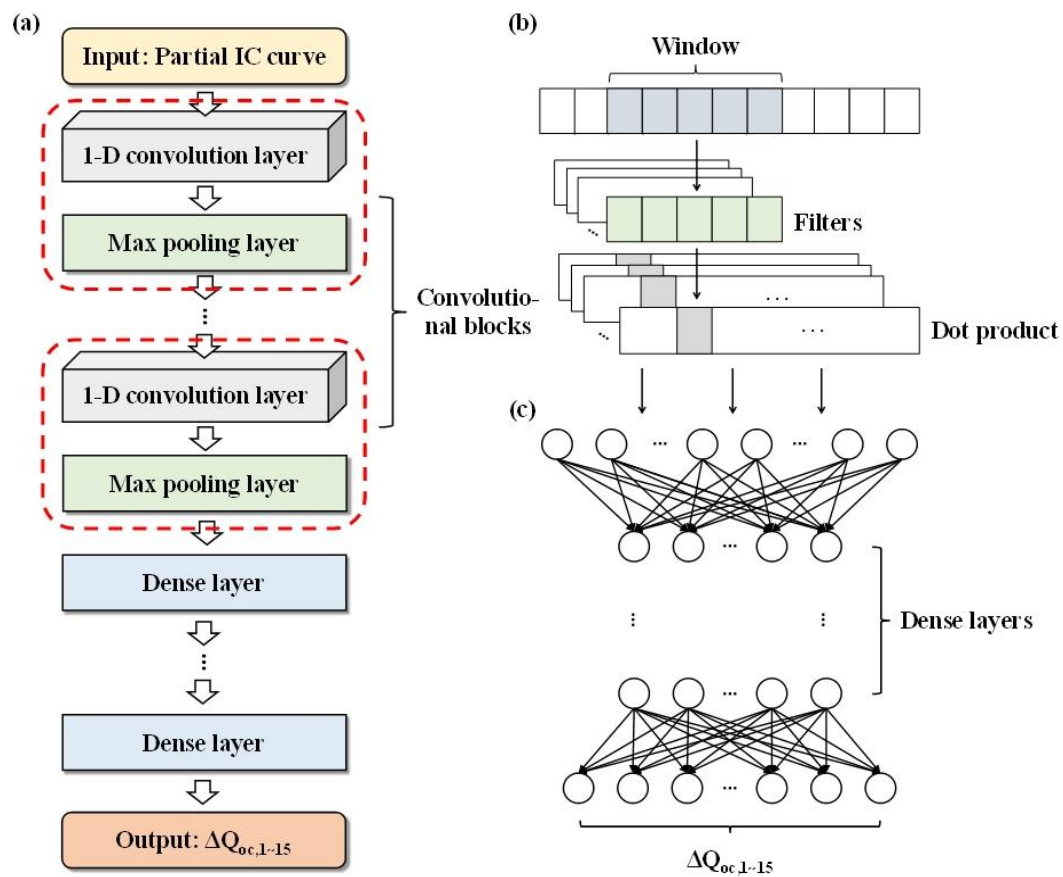

**Fig. 9.** Architecture of the proposed CNN (a); working principle of a convolutional layer (b); structure of dense layers (c).

$$\begin{cases} f_1 = \underset{\theta_{hyper},V_1,V_2}{\arg\min} \sum_{j=1}^{N_s}\sum_{i=1}^{m}\left\{\Delta Q_{FP,i} - \Delta Q_{FP,i}^{CNN}(\theta_{hyper},V_1,V_2)\right\}^2 \\ f_2 = \underset{V_1,V_2}{\arg\min}(V_2 - V_1) \end{cases} \quad (7)$$

where $m$ is the amount of the OCV feature points and $N_s$ is the amount of training data. $\theta_{hyper}$ represents the hyperparameters of the CNN. $V_1$ and $V_2$ denote the start and end voltage of partial IC segments within [3.5 V, 4 V], determining the input size of the CNN. $\Delta Q_{FP,i}$ and $\Delta Q_{FP,i}^{CNN}$ are the reference and estimated differential charge amounts of the OCV feature points, respectively.

TABLE I lists all the optimizable items with their ranges for the proposed CNN, and a non-dominated sorting genetic algorithm II (NSGA-II) is applied to minimize the two objective functions in Eq. (7) simultaneously. As a versatile optimization algorithm that finds applications in a wide range of scenarios, NSGA-II is an improved version of NSGA with enhanced diversity preservation capability and computational efficiency [40]. The working principle of NSGA-II is illustrated in Fig. 10, and its philosophy is grounded on four key strategies: non-dominated sorting, elite preserving operator, crowding distance, and selection operator.

TABLE I

OPTIMIZABLE CNN HYPERPARAMETERS AND INPUT SIZE

| Items | Searching range |
|---|---|
| CNN layer number | [1, 5] |
| Filter number | [1, 128] |
| Filter size | [2, 12] |
| Pooling size | [2, 8] |
| Stride of pooling | [1, 5] |
| Dense layer number | [1, 5] |
| Dense layer neuron | [8, 64] |
| Initial learning rate | [0.0002, 0.01] |
| Start voltage of IC segment (i.e., $V_1$) | [3.5, $V_2$] |
| End voltage of IC segment (i.e., $V_2$) | [$V_1$, 4] |

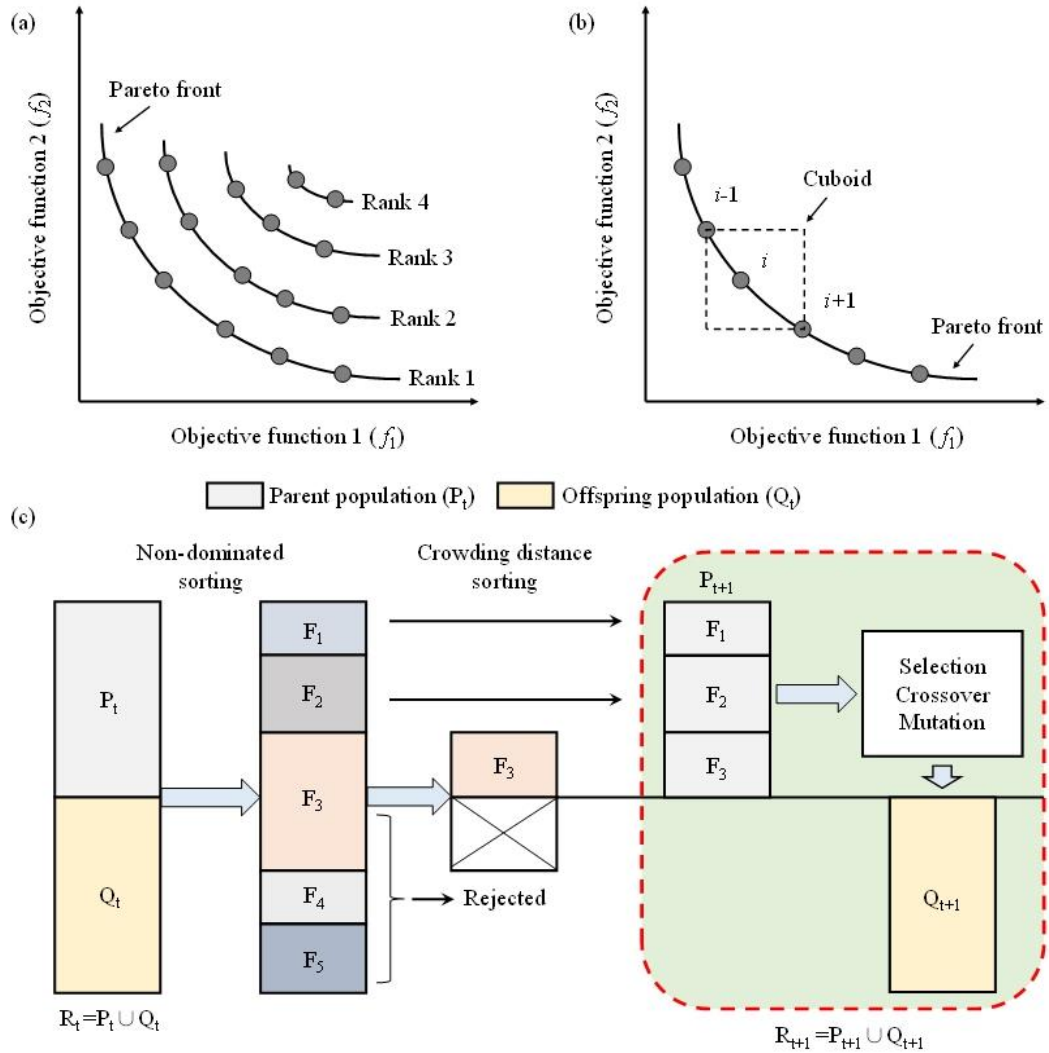

**Fig. 10.** Working principle of NSGA-II: (a) non-dominated sorting; (b) crowding distance calculation; and (3) algorithm flowchart.

*a. Non-dominated sorting*

The non-dominated sorting organizes population individuals using the concept of Pareto dominance. Let $x^1$ and $x^2$ be two feasible solutions of a multi-objective optimization function, when we say that $x^1$ dominates $x^2$ (or $x^2$ is dominated by $x^1$), it means a better solution of $x^1$ than $x^2$ if the following two conditions hold:

$$\begin{cases} f_j(x^1) \leq f_j(x^2) \text{ for all } j = \{1, 2, ..., k\} \\ f_j(x^1) < f_j(x^2) \text{ for at least one } j = \{1, 2, ..., k\} \end{cases} \quad (8)$$

where $k$ is the number of objective functions and $f_j$ denotes the $j$-th objective function. When a solution $x$ is not dominated by any other feasible solutions, it is called a Pareto optimal solution. The collection of all Pareto optimal solutions is known as a Pareto set, and the objective vector deriving from a Pareto set is defined as a Pareto front.

The process of a non-dominated sorting begins with assigning the first rank to non-dominated individuals within the initial population. These top-ranked individuals are then grouped in the first front and removed from the initial population. To proceed, the same procedure is applied to the remaining population individuals, and this iterative process continues until all the population individuals have been allocated to different fronts based on their ranks, as illustrated in Fig. 10 (a).

*b. Elite preserving operator*

The elite preserving strategy preserves the best solutions of a population and transmits them directly to the next generation. As a consequence, the non-dominated solutions identified in each generation continue to exist in subsequent generations until they are dominated by other solutions.

*c. Crowding distance*

The crowding distance is used to gauge the concentration in the vicinity of a specific solution. As shown in Fig. 10 (b), it represents the average distance of the two nearest solutions on either side, as given by

$$cd(i) = \sum_{i=1}^{k} \frac{f_j^{i+1} - f_j^{i-1}}{f_j^{\max} - f_j^{\min}} \quad (9)$$

where $f_j^i$ denotes the *i*-th individual of the *j*-th objective function. $f_j^{max}$ and $f_j^{min}$ stand for the maximum and minimum value of the *j*-th objective function among all the individuals. When comparing two solutions with differing crowding distances, the one with the greater crowding distance is regarded as being located in a less densely populated region.

  *d. Selection operator*

The next-generation population is determined through a crowded tournament selection operator, considering the rank and crowding distances of population individuals for selection. The selection rule is as follows:

  (i) When two population individuals have different ranks, the one with the superior rank is chosen for the next generation.

  (ii) In the cases where both population individuals hold the same rank, the one with the greater crowding distance is preferred for the next generation.

  To implement NSGA-II in optimizing the hyperparameters and input size of the CNN, a number of 200 initial parent population will be randomly generated from the given searching ranges. The fraction of crossover is set as 0.8, and the fraction of population on non-dominated front is set as 0.2. The algorithm terminates after going through 20 iterations, and all the computations are executed on a laptop (AMD Ryzen 5 @ 2.10 GHz, 8GB RAM).

3.3. Rapid estimation of electrode aging parameters

  Having the relative positions of the OCV feature points (i.e., $\{\Delta Q_{FP,i}^{CNN}, V_{FP,i}\}$, *i*=1,…,15) from the proposed CNN, we intend to identify the three optimal EAPs (i.e., $Q_{PE}$, $Q_{NE}$ and $Q_{offset}$) that reconstruct these points as perfectly as possible. Given a set of EAPs derives a battery OCV curve ranging from 2.7 V to 4.2 V, on which any points can be expressed as

$$V_{oc}^{EAP} = f_{PE}(x_{PE}) - f_{NE}(x_{NE}) \qquad (10)$$

with the PE and NE SOCs described by

$$\begin{cases} x_{PE} = 1 - \dfrac{Q_x}{Q_{PE}} \\ x_{NE} = \dfrac{Q_x - Q_{offset}}{Q_{NE}} \end{cases} \qquad (11)$$

where $V_{oc}^{EAP}$ denotes the EAP-derived battery OCV. $Q_x$ stands for the x-coordinate of an OCV point in the coordinate system defined in Fig. 4.

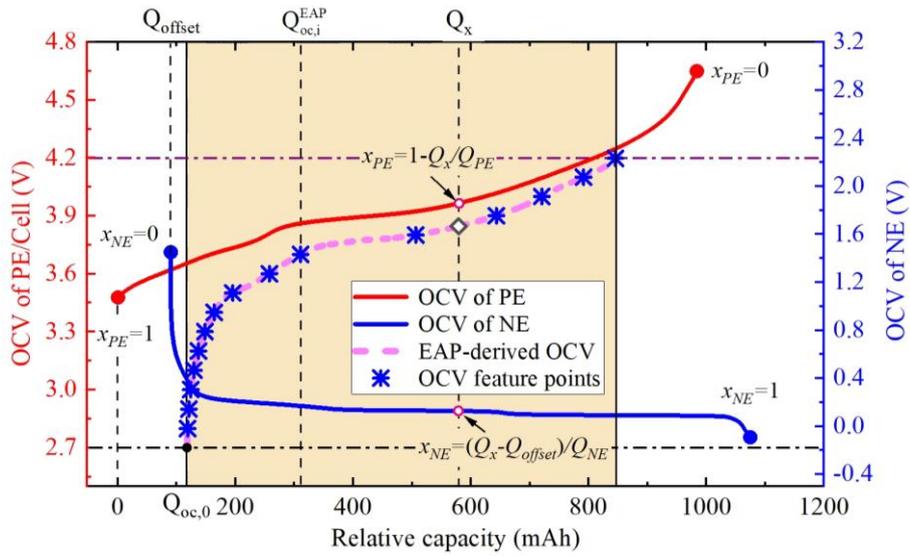

**Fig. 11.** Battery OCV reconstruction and EAPs estimation.

Let $Q_{oc,0}$ be the x-coordinate of the starting point of an EAP-derived battery OCV curve, we leverage Eqs. (10)-(11) to locate the coordinate of the $i$-th OCV feature point on this curve as

$$\begin{cases} Q_{FP,i}^{EAP} = Q_{oc,0} + \sum_{j=1}^{i} \Delta Q_{FP,i}^{CNN} \\ V_{FP,i}^{EAP} = f_{PE}(1 - Q_{FP,i}^{EAP}/Q_{PE}) \\ \qquad - f_{NE}((Q_{FP,i}^{EAP} - Q_{offset})/Q_{NE}) \end{cases} \qquad (12)$$

where $Q_{FP,i}^{EAP}$ and $V_{FP,i}^{EAP}$ denote the x and y-coordinate of the $i$-th OCV feature point, respectively, on an EAP-derived battery OCV curve. A coordinate representation of these OCV feature points allows us to determine the three optimal EAPs by minimizing the sum of the squared errors between the EAP-derived and reference feature voltages.

$$L = \arg\min_{\theta_{EAP}} \frac{1}{m} \sum_{i=1}^{m} \{V_{FP,i} - V_{FP,i}^{EAP}(\theta_{EAP}, \Delta Q_{FP,i=1,\ldots,15}^{CNN})\}^2 \qquad (13)$$

where $\theta_{EAP}$ represents the three EAPs. A hybrid particle swarm optimization (PSO)-genetic algorithm (GA) method is employed in this study to search EAPs. It harnesses the strengths of two well-known optimization algorithms, facilitating a swift convergence of candidate solutions towards the global optimum while mitigating the risk of being trapped into local minima. The main steps of the PSO-GA method are shown in Fig. 12.

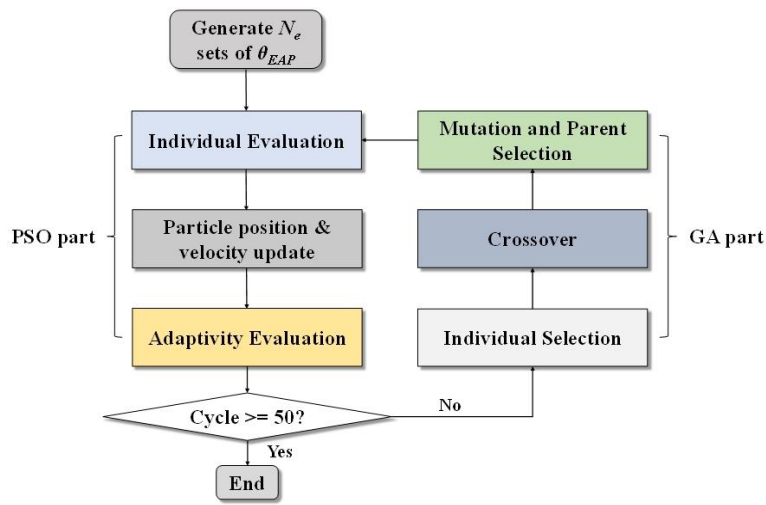

**Fig. 12.** Flowchart of hybrid PSO-GA method.

3.4. Clustering algorithm

An adAP algorithm is employed in this study to sort RBs with their respective EAPs. Compared to conventional clustering algorithms (e.g., KMC), the AP algorithm leverages the concept of 'message passing' among all samples, unlocking the necessity of pre-determining the clustering number [41]. As an enhancement of generic AP algorithm, the adAP algorithm further addresses the issues with regard to preference tuning, convergence, and oscillations during iterations, making it well-suited for sorting RBs with unknown clustering number [42].

The adAP algorithm starts with calculating the similarity matrix $S$ for the entire dataset $\Xi = \{\theta_{EAP,1}, \theta_{EAP,2}, ..., \theta_{EAP,n}\}$.

$$S(i,j) = -\|\theta_{EAP,i} - \theta_{EAP,j}\|^2 \text{ for } (i \neq j) \tag{14}$$

where $n$ denotes the total amount of RBs to be sorted. The diagonal of a $n$-by-$n$ similarity matrix $S$ is a critical input parameter, called 'preference' and denoted by $p$, which reflects the likelihood of a specific sample being a clustering center (referred to as 'exemplar' in the adAP algorithm). After obtaining the similarity matrix $S$, the responsibility and availability of two samples are evaluated by

$$\begin{cases} R(i,j) = S(i,j) - \max_{j' \neq j}\{A(i,j') + S(i,j')\} \\ R(j,j) = S(j,j) - \max_{j' \neq j}\{S(j,j')\} \end{cases} \tag{15}$$

and

$$\begin{cases} A(i,j) = \min\left\{0,\ R(j,j) + \sum_{i' \notin \{i,j\}} \max\left\{0,\ R(i',j)\right\}\right\} \\ A(j,j) = \sum_{i' \neq j} \max\left\{0,\ R(i',j)\right\} \end{cases} \quad (16)$$

where the responsibility $R(i, j)$ denotes the attraction messages sent from the $i$-th sample to the $j$-th sample, quantifying the suitability of the latter to serve as an exemplar for the former in the presence of the other potential exemplars within the dataset. In response to an attraction message, the availability $A(i, j)$ reflects the appropriateness of the $j$-th sample being the exemplar for the $i$-th sample. The responsibility $R(i, j)$ and availability $A(i, j)$ will be iteratively updated by Eq. (17) based on a damping factor $\lambda$.

$$\begin{cases} R_w(i,j) = (1-\lambda)R_w(i,j) + \lambda R_{w-1}(i,j) \\ A_w(i,j) = (1-\lambda)A_w(i,j) + \lambda A_{w-1}(i,j) \end{cases} \quad (17)$$

where the subscript $w$ denotes the iteration index. The damping factor $\lambda$ is initialized as 0.5.

The adAP algorithm determines the clustering number for a dataset by the input preference $p$; however, it remains unknown for which $p$ corresponds to the optimal clustering number. Generally, a $p$ close to the minimum similarity produces fewer clusters while a $p$ close to or larger than the maximum similarity produces more clusters. An adaptive $p$-scanning strategy is, therefore, developed to overcome this issue. In this strategy, the preference $p$ is updated in association with the current $K$ exemplars, and it decreases continually by Eq. (18) at every iteration if the number of exemplars converges to $K$ (i.e., all the exemplars remain unchanged over a specified monitoring window).

$$\begin{cases} p = p - c \times p_{step} \\ p_{step} = 0.01 \times p_{median} / q \\ q = 0.1 \times \sqrt{K} + 50 \end{cases} \quad (18)$$

where $c$ is a counter for the convergence of $K$ exemplars at every iteration and $p_{median}$ denotes the median of the similarity. When implementing the adaptive $p$-scanning strategy, an increment of 0.05 will be imposed on the damping factor $\lambda$ in case of oscillation occurring within a monitoring window.

The exemplars are identified by Eq. (19), which gives a criterion matrix based on the summation of the responsibility $R(i, j)$ and availability $A(i, j)$.

$$C_r(i, j) = R(i, j) + A(i, j) \quad (19)$$

An exemplar refers to the element with the maximum criterion value $C_r(i, j)$, and the row elements belonging to the same exemplar are assigned into one cluster. The iteration terminates when the number of exemplars reaches 2, establishing the lowest preference threshold.

Assuming that all the samples have been assigned into clusters $C_i$ ($i=1,\ldots,K$), we assess the silhouette coefficient $Sil$ by Eq. (20) for each sample, and the largest overall average silhouette coefficient gives the best clustering performance and optimal clustering number.

$$Sil(t) = \frac{b(t) - a(t)}{\max\{a(t), b(t)\}} \quad (20)$$

where $a(t)$ is defined as the average dissimilarity of the $t$-th sample of $C_j$ to all the other samples within $C_j$. Let $d(t, C_i)$ be the average dissimilarity of the $t$-th sample of $C_j$ to all the samples in another cluster $C_i$, then $b(t)$ is defined as the minimum value

26of $d(t,C_i)$, namely $b(t) = \min\{d(t,C_i)\}$, ($i \neq j$, $i=1,\ldots,K$). It is worth noting that the dissimilarity among samples is evaluated by the Euclidean distance in this study.

**4. Results and discussions**

4.1. Validation of CNN models

NSGA-II makes different trade-offs between model precision and input size of the proposed CNN. Fig. 13 presents the non-dominated Pareto front of the optimization results. Three candidates are selected on this non-dominated Pareto front as the final solutions, labeled as CNN #1~3. As the primary criterion of selection, we prioritize Objective 1 over Objective 2, that is, the proposed CNN must first and foremost meet the requirement of high model accuracy, and it is only after satisfying this condition that we consider minimizing the input size. Consequently, the first model, i.e., CNN #1, is chosen because of its highest accuracy, although it demands the lengthiest partial IC segment of around 300 data points in the non-dominated set. Compared to the first model, the second model, i.e., CNN #2, significantly reduces the input size to around 200 data points with slight performance deterioration, showing promising feasibility and efficiency in practical applications. The third models, i.e., CNN #3, further decreases the input size to only 120 data points yet sacrificing model performance to a certain extent. It is worth noting that we sample the IC curve in steps of 1 mV. In other word, an input size of 120 data points, for instance, indicates a voltage interval of only 0.12 V or so (corresponding to a 1-C charging time of around 13 min), as visualized in Fig. 14. TABLE II lists the detailed hyperparameters and input size of CNN #1~3. The performances of these well-optimized and trained CNNs in relocating the relative positions of the OCV feature points are validated based on the degradation data from Cell #4~8.

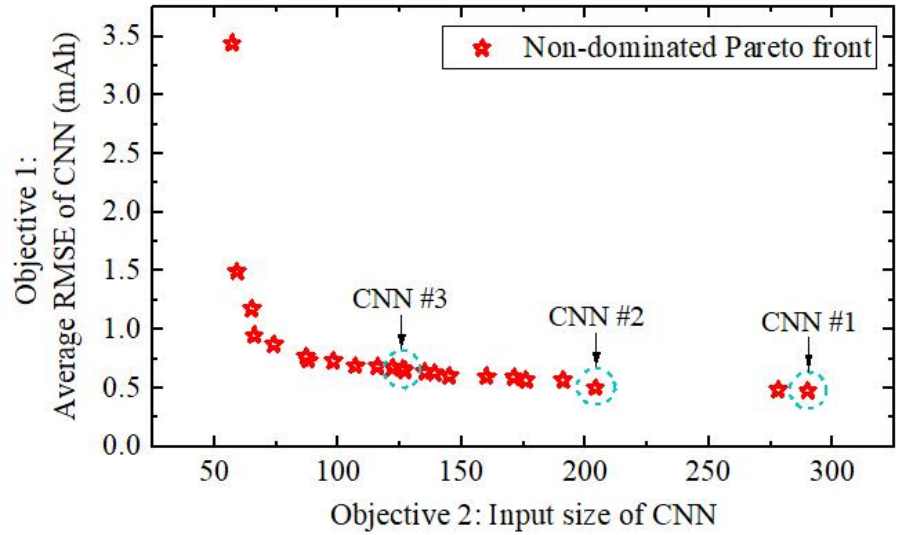

**Fig. 13.** Non-dominated Pareto front of the optimization results for CNN hyperparameters and input size.

TABLE II

OPTIMIZATION RESULTS OF CNN HYPERPARAMETERS AND INPUT SIZE

| Items | CNN #1 | CNN #2 | CNN #3 |
| --- | --- | --- | --- |
| CNN layer number | 2 | 1 | 3 |
| Filter number | 13 | 9 | 10 |
| Filter size | 26 | 29 | 31 |
| Pooling size | 3 | 3 | 3 |
| Stride of pooling | 1 | 1 | 1 |
| Dense layer number | 1 | 1 | 1 |
| Dense layer neuron | 45 | 49 | 39 |
| Initial learning rate | 0.0040 | 0.0032 | 0.0048 |
| Start voltage of IC segment (i.e., $V_1$) | 3.601 | 3.665 | 3.695 |
| End voltage of IC segment (i.e., $V_2$) | 3.891 | 3.869 | 3.822 |
| Average RMSE (mAh) | 0.6922 | 0.7141 | 0.8124 |

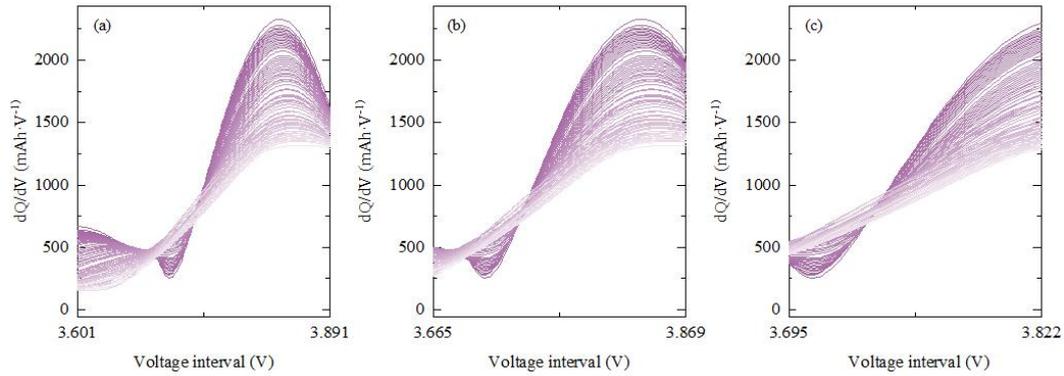

**Fig. 14.** Partial IC segments as optimized input sequences for: (a) CNN #1; (b) CNN #2; and (c) CNN #3.

Fig. 15-Fig. 17 show the parity plots of the comparisons between the estimated and reference differential charge amounts (i.e., $\Delta Q_{FP,1} \sim \Delta Q_{FP,15}$) for Cell #4~8 based on CNN #1~3, respectively. It can be seen that almost all the estimations fit the reference lines for the test samples, and only a few outliers appear with deviations of several milli ampere-hours. Compared to the entire available capacity in the order of hundreds of milli ampere-hours, those outliers will have little influences on EAPs estimation and battery sorting.

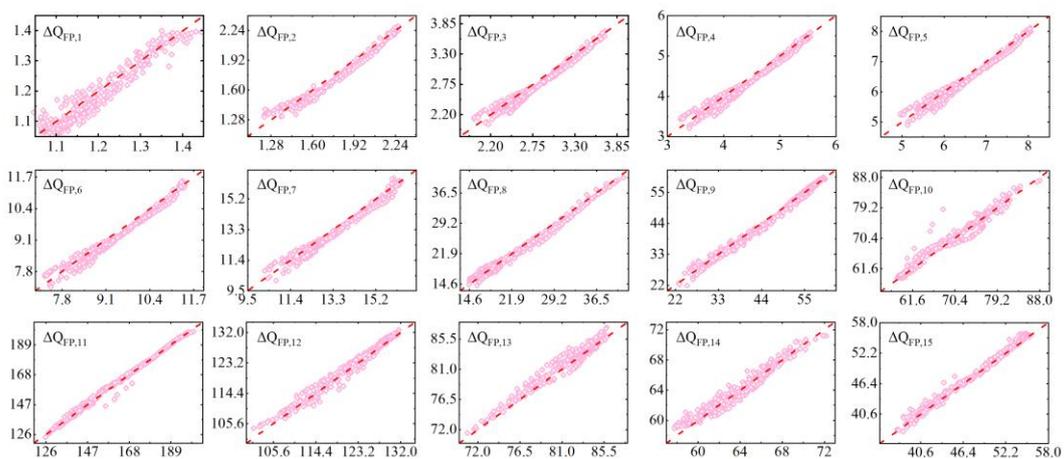

**Fig. 15.** Parity plots showing the comparisons of the estimated and reference differential charge amounts (i.e., $\Delta Q_{FP,1} \sim \Delta Q_{FP,15}$) for Cell #4~8 based on CNN #1.

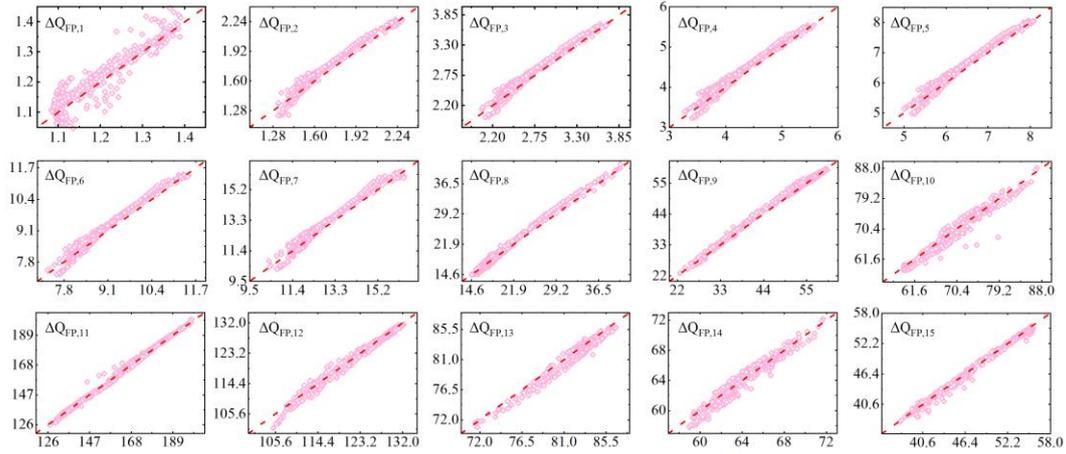

**Fig. 16.** Parity plots showing the comparisons of the estimated and reference differential charge amounts (i.e., $\Delta Q_{FP,1} \sim \Delta Q_{FP,15}$) for Cell #4~8 based on CNN #2.

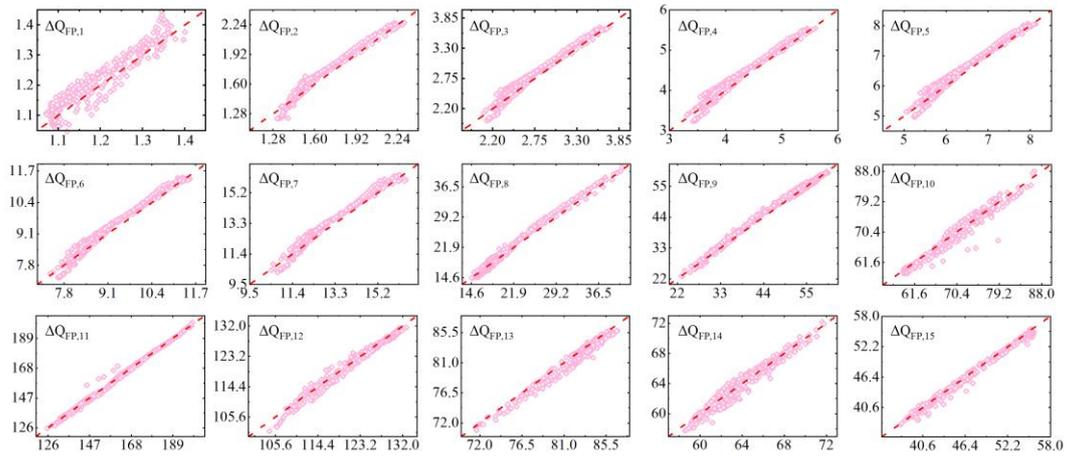

**Fig. 17.** Parity plots showing the comparisons of the estimated and reference differential charge amounts (i.e., $\Delta Q_{FP,1} \sim \Delta Q_{FP,15}$) for Cell #4~8 based on CNN #3.

4.2. Validation of EAPs estimation

The EAPs estimation of Cell #4~8 is implemented by reconstructing battery OCV curves at specific aging levels and minimizing their deviations to the estimated relative positions of OCV feature points. The upper and lower cut-off voltages delineate the usable range of a reconstructed battery OCV curve as well as the associated available capacity.

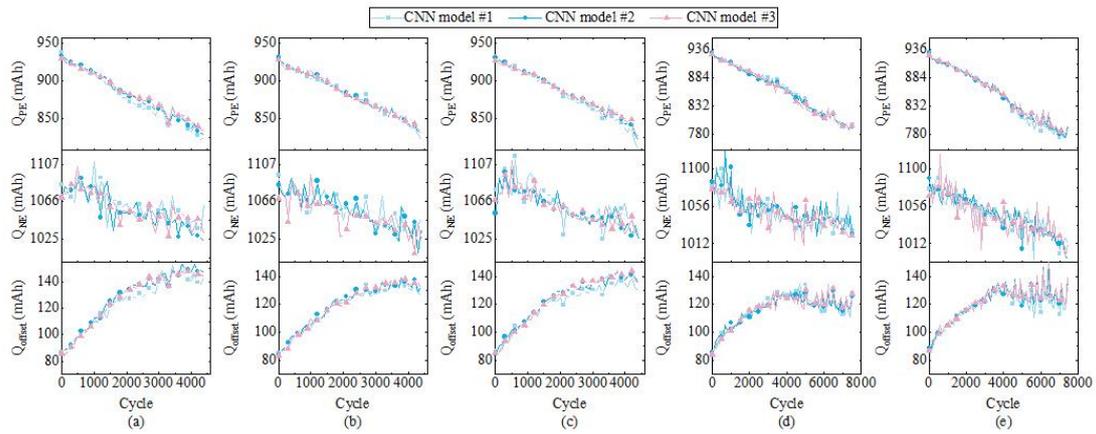

**Fig. 18.** EAPs estimation results based on CNN #1~3 for: (a) Cell #4; (b) Cell #5; (c) Cell #6; (d) Cell #7; and (e) Cell #8.

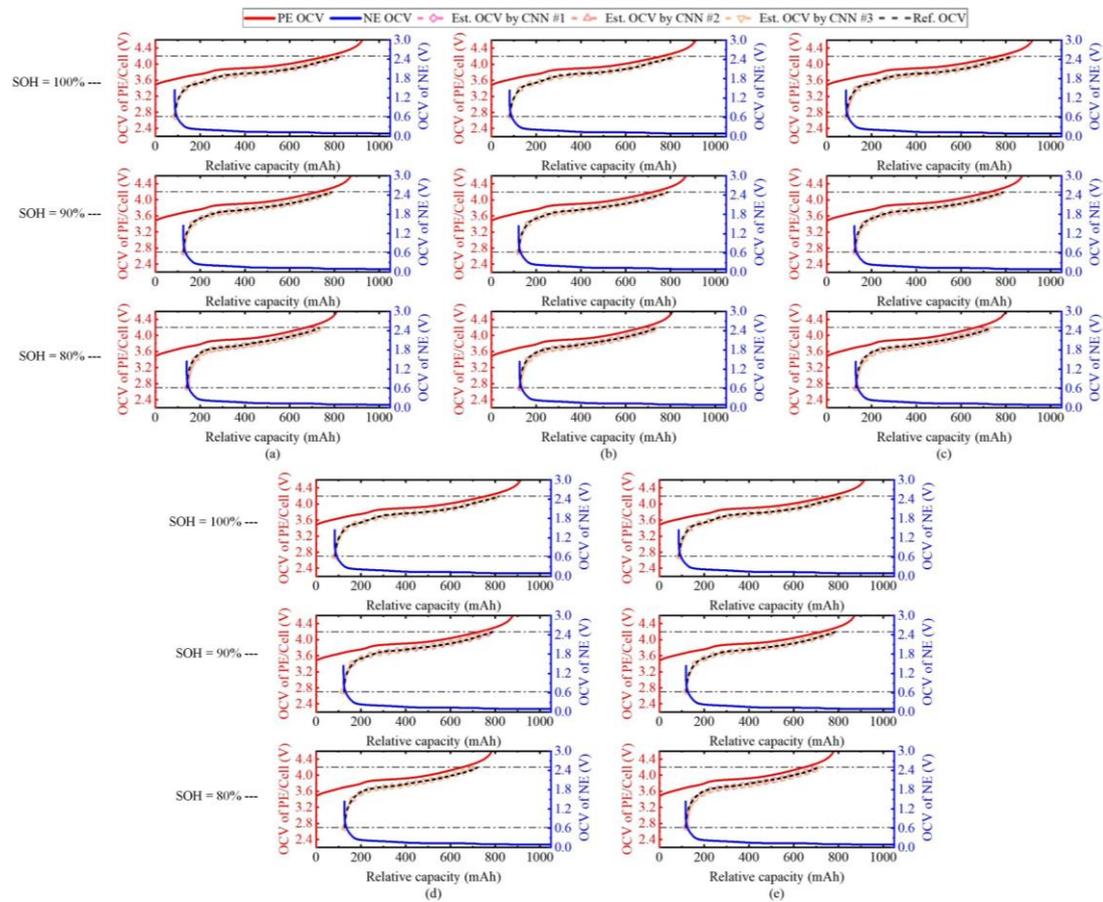

**Fig. 19.** Battery OCV reconstruction results by fitting 15 OCV feature points: (a) Cell #4; (b) Cell #5; (c) Cell #6; (d) Cell #7; and (e) Cell #8.

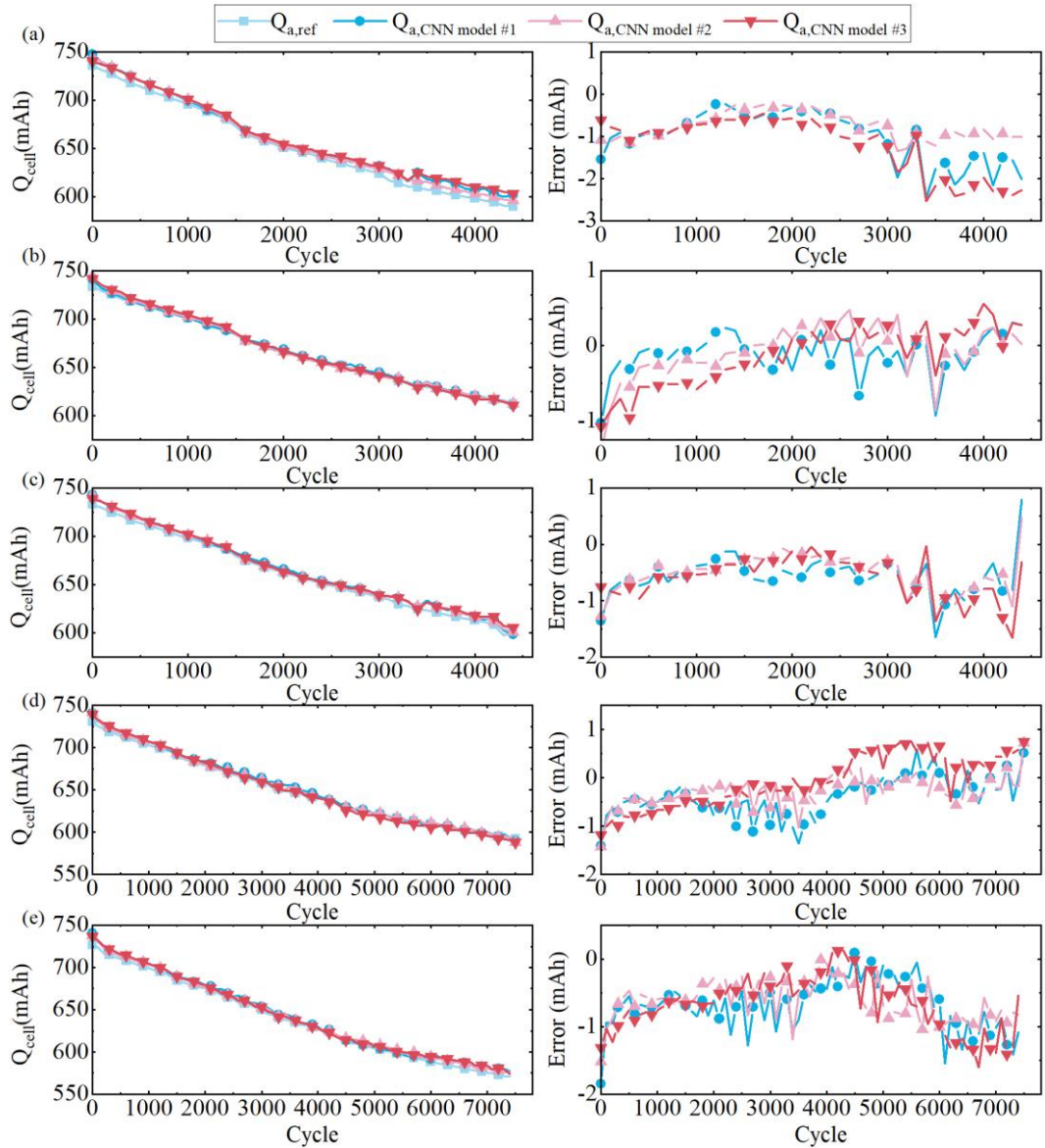

**Fig. 20.** Capacity estimation results based on CNN #1~3 for: (a) Cell #4; (b) Cell #5; (c) Cell #6; (d) Cell #7; and (e) Cell #8.

Fig. 18 shows the estimated three EAPs based on CNN #1~3 over the whole battery lifetime of Cell #4~8. As we can see that CNN #1~3 outcome the EAPs estimation in high consistency, where all these variation trajectories agree well with the aging mechanism analysis given in Section III. A, indicating a highly correlated relationship with battery degradation. Having the EAPs estimation of each cell, we also validate their performances in battery OCV reconstruction and capacity estimation as shown in Fig. 19 and 20, respectively. According to these results, the reconstructed battery OCV curves based on CNN #1~3 exhibit significant overlaps with the reference curves at various aging levels, and the overall mean absolute error (MAE) and RMSE for capacity estimation of Cell #4~8 (see TABLE III) are below 0.7% and 0.8%, respectively, confirming the high fidelity of the EAPs estimation.

TABLE III

ESTIMATION ERRORS OF BATTERY AVAILABLE CAPACITY FOR CELL #4~8

| Models | Estimation error (%) | |
|---|---|---|
| | MAE | RMSE |
| CNN #1 | 0.49 | 0.60 |
| CNN #2 | 0.59 | 0.74 |
| CNN #3 | 0.63 | 0.79 |

4.3. Validation of battery sorting

In this study, a total of 150 cycles of battery data is randomly picked up from different aging levels of Cell #4~8 to simulate the testing results of 150 RBs after being dismantled from a battery pack in EVs. The available capacities of these RBs simulated by the degradation data from Cell #4~8 are presented in Fig. 21, which range from 735.98 mAh (99.34% of the nominal capacity) to 571.89 mAh (77.15% of the nominal capacity) with a standard deviation (SD) of 44.29 mAh, amounting to 5.98% of the nominal capacity. In this way, we create the simulation dataset with remarkable in-pack heterogeneity from both individual aspect and degradation aspect, guaranteeing the convincingness of the validation results.

Fig. 22 visualizes the 3-D clustering of the proposed method via the EAPs estimated based on CNN #1~3, and the associated optimal clustering numbers are 8, 9 and 10, respectively. Fig. 23-Fig. *25* show the EAPs and available capacities of RBs in different clusters. It can be seen that the intra-cluster heterogeneity, including both EAPs and available capacities, are significantly mitigated after sorting, indicating the effectiveness of the proposed method.

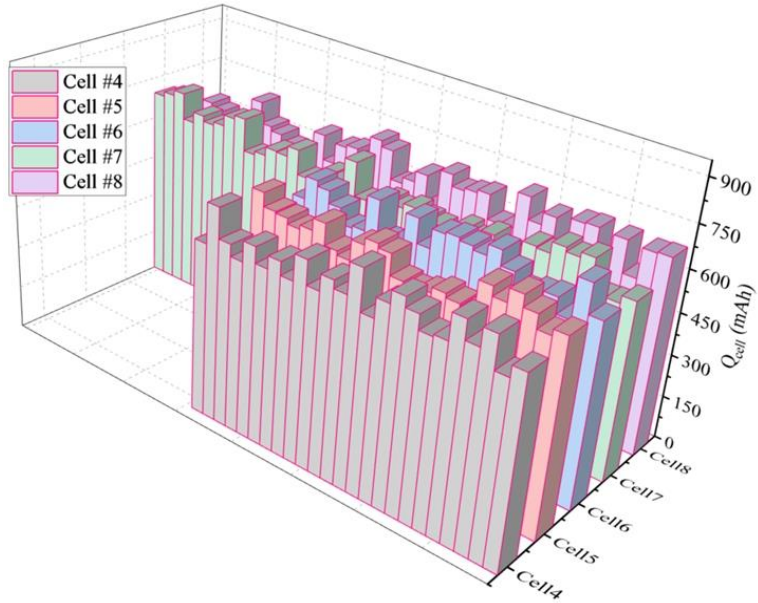

**Fig. 21.** Battery data selection at different aging levels of Cell #4~8 for validation of the proposed method in battery sorting.

TABLE IV

AVERAGE STANDARD DEVIATION OF BATTERY SORTING RESULTS VIA THE EAPs ESTIMATED BASED ON CNN #1

| Methods | Average SD (mAh) | | | |
| --- | --- | --- | --- | --- |
| | $Q_{cell}$ | $Q_{PE}$ | $Q_{NE}$ | $Q_{offset}$ |
| Proposed method (K=8) | 9.06 | 6.48 | 10.04 | 5.17 |
| KMC (K=8) | 9.75 | 12.13 | 11.32 | 5.21 |
| FCMC (K=8) | 9.29 | 11.95 | 11.86 | 5.05 |
| AHC (K=8) | 10.04 | 11.39 | 10.88 | 5.55 |

To validate the superiority of the proposed method in battery sorting, we take the conventional capacity based sorting methods as benchmarks, which utilize two IC peaks and one valley (see Fig. 8) as health indices and construct a three-layer fully-connected neural network to estimate battery available capacity, followed by three commonly used algorithms for clustering (i.e., KMC algorithm, fuzzy C-means clustering (FCMC) algorithm, and agglomerative hierarchical clustering (AHC) algorithm). For quantitative analysis, we also implement the benchmark methods by initializing the clustering numbers as 8, 9 and 10 to make fair comparisons with the proposed method.

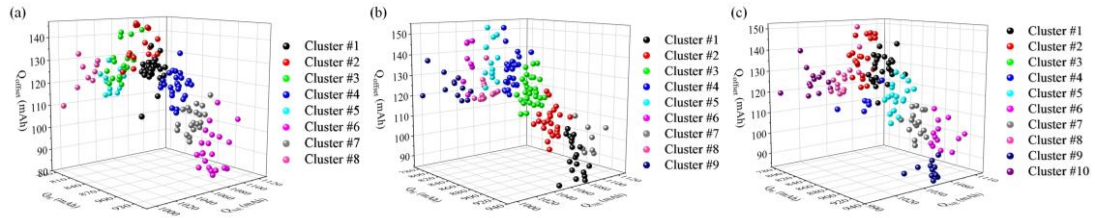

**Fig. 22.** 3-D representations of clustering results via the EAPs estimated based on: (a) CNN #1; (b) CNN #2; and (c) CNN #3.

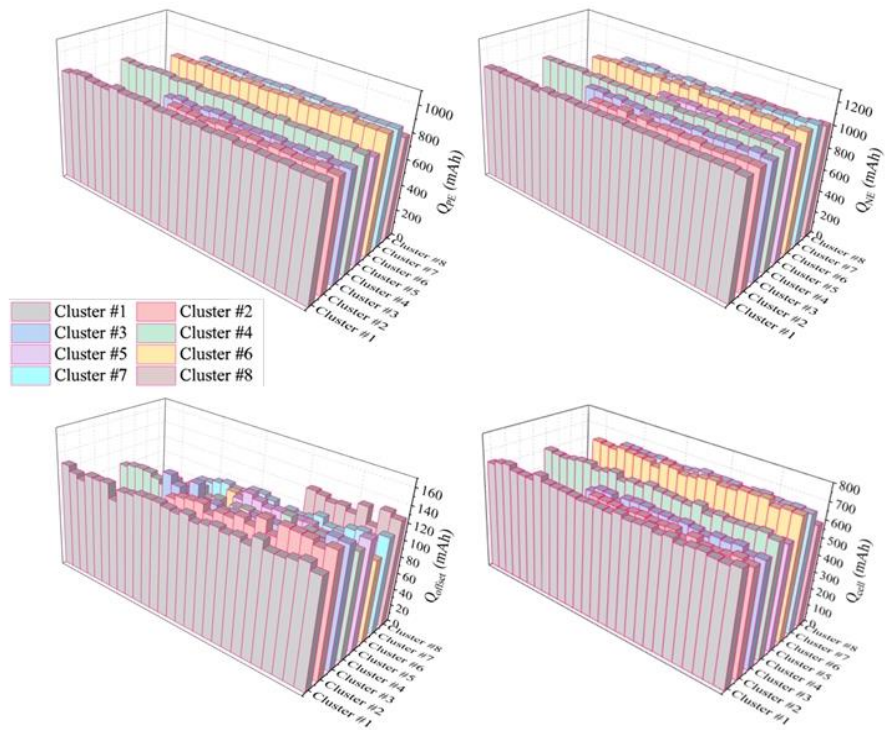

**Fig. 23.** Battery sorting results via the EAPs estimated based on CNN #1.

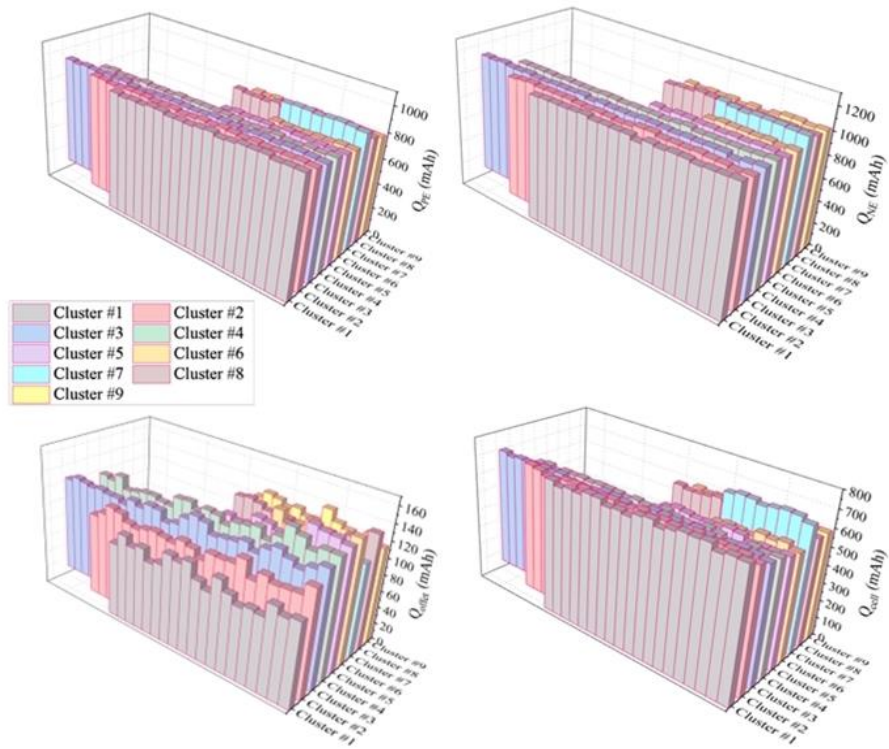

**Fig. 24.** Battery sorting results via the EAPs estimated based on CNN #2.

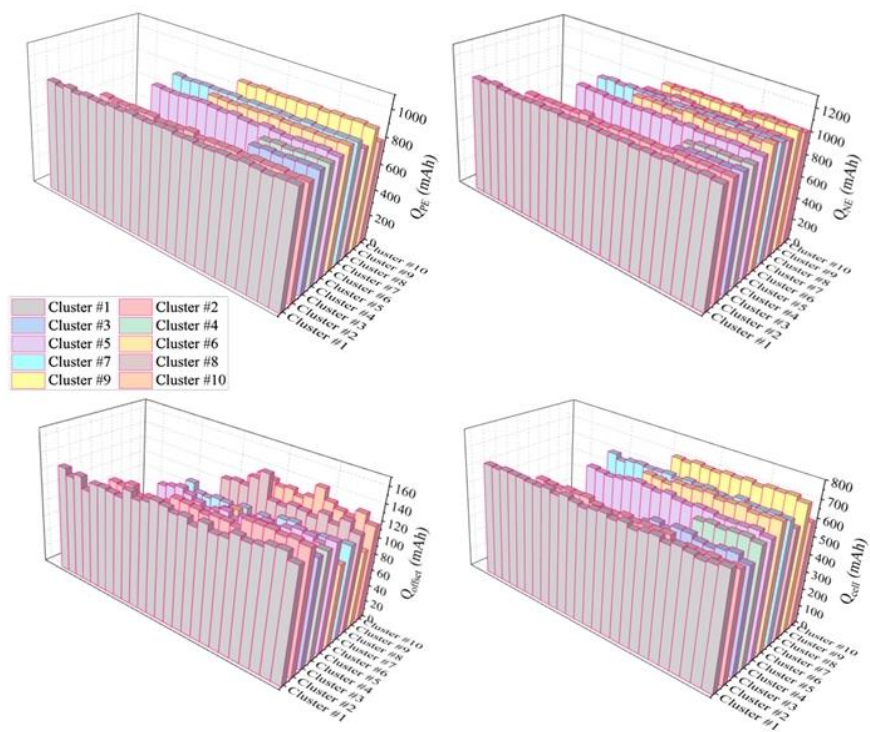

**Fig. 25.** Battery sorting results via the EAPs estimated based on CNN #3.

TABLE IV-TABLE VI list the numerical results of different methods. As can be observed from the statistics, the proposed method outperforms the other benchmark methods in sorting RBs with similar degradation patterns of PE and NE, which largely improves the homogeneities of $Q_{PE}$ and $Q_{NE}$ by reducing their average SDs to around 6 mAh and 9 mAh, respectively. Comparatively, the heterogeneities of EAPs are somehow noticeable after sorting by the benchmark methods, although the average SDs of available capacity are close to the proposed method in some cases. This discloses the fact that the RBs with similar available capacities can feature different degradation patterns at an electrode level, highlighting the significance of battery sorting from a microscopic scale.

TABLE V

AVERAGE STANDARD DEVIATION OF BATTERY SORTING RESULTS VIA THE EAPS ESTIMATED BASED ON CNN #2

| Methods | Average SD (mAh) | | | |
| --- | --- | --- | --- | --- |
| | $Q_{cell}$ | $Q_{PE}$ | $Q_{NE}$ | $Q_{offset}$ |
| Proposed method (K=9) | 8.06 | 6.35 | 9.02 | 4.79 |
| KMC (K=9) | 8.91 | 11.02 | 11.09 | 4.89 |
| FCMC (K=9) | 8.64 | 11.11 | 11.05 | 4.97 |
| AHC (K=9) | 9.78 | 11.77 | 10.84 | 4.66 |

TABLE VI

AVERAGE STANDARD DEVIATION OF BATTERY SORTING RESULTS VIA THE EAPS ESTIMATED BASED ON CNN #3

| Methods | Average SD (mAh) | | | |
| --- | --- | --- | --- | --- |
| | $Q_{cell}$ | $Q_{PE}$ | $Q_{NE}$ | $Q_{offset}$ |
| Proposed method (K=10) | 8.57 | 6.22 | 8.72 | 4.65 |
| KMC (K=10) | 8.33 | 10.51 | 11.38 | 4.76 |
| FCMC (K=10) | 8.74 | 12.27 | 11.01 | 4.81 |
| AHC (K=10) | 8.84 | 10.47 | 10.13 | 5.37 |

## 5. Conclusions

In this study, we developed an electrode aging assessment approach for accurate and efficient sorting of RBs. Unlike conventional sorting methods based solely on capacity, the proposed method allows for the characterization of different degradation patterns at an electrode level, minimizes the need for CC charging data, and supports module/pack-level tests for simultaneous processing of high volumes of RBs. The battery degradation data of different cells at various aging levels (77.15% to 99.34% of the nominal capacity) are leveraged to simulate 150 RBs dismantled from a battery pack in EVs. The validation results indicate the promising performance of the proposed method, where all the RBs show great homogeneity in their respective groups with SDs of around 6 mAh in $Q_{PE}$, 9 mAh in $Q_{NE}$, 5 mAh in $Q_{offset}$, and 8 mh in $Q_{cell}$. This makes a significant improvement in contrast to capacity based benchmark methods.

**Declaration of competing interest**

The authors declare that they have no known competing financial interests or personal relationships that could have appeared to influence the work reported in this paper.

**Acknowledgement**

This research is supported by an Australian government research training program scholarship offered to the first author of this study. The authors would also like to thank Dr Birkl and Prof Howey for providing the battery dataset.